# ACOUSTICS OF FINITE-APERTURE VORTEX BEAMS

## Farid G. MITRI[1,*]

[1] Chevron – Area 52 Technology, Santa Fe, 87508, United States

*Corresponding author: mitri@chevron.com, fmitri@gmail.com

**Abstract**

A method based on the Rayleigh-Sommerfeld surface integral is provided, which makes it feasible to rigorously model, evaluate and compute the acoustic scattering and other mechanical effects of finite-aperture vortex beams such as the acoustic radiation force and torque on a viscoelastic sphere in various applications in acoustic tweezers and microfluidics, particle entrapment, manipulation and rotation. Partial-wave series expansions are derived for the incident field of acoustic spiraling (vortex) beams, comprising high-order Bessel and Bessel-Gauss beams.

## 1 Introduction

Considering the fact that practically every acoustic source (except a point source radiating omnidirectional waves) produces a finite beam, it is necessary to devise a method which enables modelling the incident pressure field for applications in beam-forming design, and other areas including imaging, scattering and dynamic effects caused by acoustic beams emanating from a finite aperture. Previous developments considered the cases of finite piston sources with uniform vibration [1], or clamped edges, Gaussian [2, 3] and others [4-7], to the zero-order Bessel beam [8, 9], for which both the axial scattering [10] and radiation force [11] on a sphere have been recently evaluated. The aim here is to extend the previous developments and provide partial-wave series expansions (PWSEs) for the acoustic pressure (scalar) field of vortex beams. In the present analysis, high-order Bessel and Bessel-Gauss beams are considered, however, the analysis can be directly extended to other types of vortex beams, such as the Laguerre-Gauss beam and other high-order Gaussian beams. The PWSEs are useful for further development related to the scattering and radiation forces by such beams, which will be the subject of future investigations.

## 2 Method

Consider a finite circular transducer with a surface of radius $b$ in a non-viscous fluid medium. A spherical coordinates system $(r, \theta, \phi)$ is chosen to coincide with the center $O$ of the Cartesian coordinates system $(x, y, z)$ (Fig. 1). The coordinates systems are situated at an axial distance $r_0$ from the center of the circular transducer. The description of the incident acoustic field produced by the finite source, and represented by its scalar velocity potential field $\Phi_i$, is obtained from the Rayleigh-Sommerfeld integral as [12],

$$\Phi_i = \frac{1}{2\pi} \iint_{S_r} \frac{v e^{i(kR-\omega t)}}{R} dS_r, \qquad (1)$$

where $R$ is the distance from the observation point to the finite source of circular surface $S_r$, the parameter $k$ is the wave number of the acoustic radiation, $\omega$ is the angular frequency, and $v$ is the normal velocity at $z = 0$. In the case of a uniform vibrational surface, $v$ is unitary [3-5, 7, 13], but takes particular values when the radiator is simply supported [3, 4, 6], clamped [3, 4], described by a rotating (propeller) [14] or ring [15] source, apodized by a Gaussian distribution [4, 16], or excited according to its radially-symmetric vibrational modes describing zero-order Bessel beams [8-11].

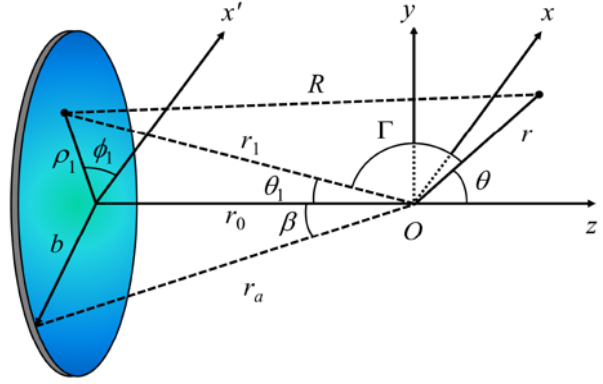

*Figure 1* Geometry of the problem used in the derivation of the incident acoustic fields of finite vortex beams.

For the purpose of the present study, apodizations of the normal velocity at $z = 0$, proportional to $\exp(im\phi)$ describing a vortex beam are considered. Current related research is investigating the scattering [17-26], radiation forces [27-30] and torques [31-33] on a spherical particle, in which infinite high-order Bessel beams were considered. However, acoustical vortex beams which carry an orbital angular momentum and emanate from a finite aperture are practically used in particle manipulation and rotation applications [34-38], and the present investigation could be potentially useful in modelling the mechanical effects (i.e. scattering, radiation forces and torques) of finite sound beams on a sphere, as opposed to waves of infinite extent.

First, the analysis treats the case of a high-order Bessel-Gaussian beam which reduces under a special limit to the case of a high-order Bessel vortex beam.

## 2.1 High-order Bessel-Gauss vortex beam

Consider an acoustic vortex for which the normal velocity is expressed as,

$$v|_{z=0} = V_0 J_m(k_\rho \rho_1) e^{-(\rho_1^2/w_0^2)} e^{im\phi_1}, \quad (2)$$

where $V_0$ is the velocity amplitude, $\rho_1$ is the distance from the center of the radiator to a point on its flat surface, $k_\rho = k \sin\beta$, with $\beta$ corresponding to the half-cone angle of the beam, $w_0$ is the beam's waist, the parameter $m \neq 0$ is a positive or negative integer describing the order of the beam, and $J_m(.)$ is the cylindrical Bessel function of the first kind of order $m$, and $\phi_1$ is the azimuthal angle at the surface of the finite source (Fig. 1).

Suppressing the time dependence $e^{-i\omega t}$ for convenience, and using the addition theorem for the spherical functions, i.e. (10.1.45) and (10.1.46) in [39] such that $r \leq r_1$, Eq.(1) is expressed as,

$$\Phi_i = \frac{ikV_0}{2\pi} \sum_{n=0}^{\infty} (2n+1) j_n(kr) \\ \times \iint_{S_r} J_m(k_\rho \rho_1) e^{-(\rho_1^2/w_0^2)} h_n^{(1)}(kr_1) e^{im\phi_1} P_n(\cos\Gamma) dS_r, \quad (3)$$

where $j_n(\cdot)$ and $h_n^{(1)}(\cdot)$ are the spherical Bessel and Hankel functions of the first kind, $P_n(\cdot)$ are the Legendre functions, and the differential surface $dS_r = \rho_1 d\rho_1 d\phi_1 = r_1 dr_1 d\phi_1$, since $r_1^2 = \rho_1^2 + r_0^2$.

Making use of the addition theorem for the Legendre functions [40, 41] using the definition of the angles as given in Fig. 1, $P_n(\cos\Gamma)$ can be expressed as ((3.19), p. 65 in [42]),

$$P_n(\cos\Gamma) = \sum_{\ell=0}^{n} (2-\delta_{\ell,0}) \frac{(n-\ell)!}{(n+\ell)!} \\ \times P_n^\ell(\cos(\pi-\theta)) P_n^\ell(\cos\theta_1) \cos\ell(\phi-\phi_1), \quad (4)$$

where $\delta_{i,j}$ is the Kronecker delta function, and $P_n^\ell(\cdot)$ are the associated Legendre functions. Integrating both sides of Eq.(4) with respect to $\phi_1$ using the property of the following integral,

$$\int_0^{2\pi} e^{im\phi_1} \cos\ell(\phi-\phi_1) d\phi_1 = \begin{cases} 0, & \ell \neq m \\ 2\pi, & \ell = m = 0 \\ \pi e^{im\phi}, & \ell = m \neq 0 \end{cases} \quad (5)$$

gives for $\ell = m \neq 0$ after substituting Eq.(4) and Eq.(5) into Eq.(3), the incident velocity potential as,

$$\Phi_i = \frac{V_0 i e^{im\phi}}{k} \\ \times \sum_{n=|m|}^{\infty} \Lambda_m (-1)^{n+m} (2n+1) j_n(kr) \frac{(n-m)!}{(n+m)!} P_n^m(\cos\theta), \quad (6)$$

where,

$$\Lambda_m = \int_{kr_0}^{kr_a} (kr_1) J_m\left(k_\rho \sqrt{r_1^2 - r_0^2}\right) e^{-\left[(r_1^2 - r_0^2)/w_0^2\right]} h_n^{(1)}(kr_1) P_n^m\left(\frac{r_0}{r_1}\right) d(kr_1). \quad (7)$$

Eq.(6) represents the incident velocity potential for an acoustic high-order Bessel-Gauss vortex beam beam with its axis of wave propagation coinciding with the z-axis. The integral in Eq.(7) can be evaluated numerically as no closed-form solution is available yet. If $m = 0$, the field corresponds to a circular disk with uniform vibration.

## 2.2 High-order Bessel vortex beam

It can be easily noticed that for wide beams, the beam's waist $w_0 \to \infty$. Thus, Eq.(7) becomes,

$$\Lambda_m|_{w_0 \to \infty} = \int_{kr_0}^{kr_a} (kr_1) J_m\left(k_\rho \sqrt{r_1^2 - r_0^2}\right) h_n^{(1)}(kr_1) P_n^m\left(\frac{r_0}{r_1}\right) d(kr_1), \quad (8)$$

and the corresponding incident velocity potential of a high-order Bessel vortex beam can be obtained by inserting Eq.(8) into Eq.(6).

## 3 Discussion and conclusion

In this work, a method based on the Rayleigh-Sommerfeld surface integral is provided, which makes it feasible to rigorously obtain partial-wave series expansions for the incident field of acoustic spiraling (vortex) beams, comprising high-order Bessel and Bessel-Gauss beams. Those expressions are useful in the study of the axial and arbitrary scattering, radiation force and torque of finite-aperture vortex beams on a viscoelastic sphere in various applications in acoustic tweezers and microfluidics, particle entrapment, manipulation and rotation. From the expressions of the incident velocity potential (or pressure) fields, the beam-shape coefficients can be obtained and used to advantage in the numerical evaluation of the scattering and other mechanical effects of acoustical beams, especially in situations where the sphere is shifted off the beam's axis.